\documentclass[prl,aps,twocolumn,superscriptaddress,showpacs,preprintnumbers,asmath,amssymb]{revtex4-1}

\usepackage{amsfonts}
\usepackage{amssymb}
\usepackage{color}
\usepackage{graphicx}
\graphicspath{{eps+pdf+jpg/}}
\usepackage{ulem}
\usepackage{float}
\usepackage{here}
\usepackage{bm}
\usepackage{amstext}
\usepackage{amsfonts}
\usepackage{cancel}
\usepackage{dcolumn} 
\usepackage[colorlinks=true,urlcolor={blue},citecolor={blue}]{hyperref}

\begin{document}

\pdfoutput=1

\title{Lattice dynamics and phase stability of rhombohedral antimony under high pressure}
\author{A. Minelli}
\affiliation{European Synchrotron Radiation Facility, BP 220, F-38043 Grenoble Cedex, France}
\author{S. M. Souliou}
\affiliation{European Synchrotron Radiation Facility, BP 220, F-38043 Grenoble Cedex, France}
\affiliation{Institute for Solid State Physics, Karlsruhe Institute of Technology, 76131 Karlsruhe, Germany}
\author{T. Nguyen-Thanh}
\affiliation{European Synchrotron Radiation Facility, BP 220, F-38043 Grenoble Cedex, France}
\author{A. H. Romero}
\affiliation{Department of Physics and Astronomy, West Virginia University, WV-26506-6315, Morgantown, USA}
\affiliation{Facultad de Ingenier\'ia-BUAP, Apartado Postal J-39, Puebla, Pue. 72570, Mexico}
\author{J. Serrano}
\affiliation{Yachay Tech University, School of Physical Sciences and Nanotechnology, Urcuqui 100119, Ecuador}
\author{W. Ibarra Hernandez}
\affiliation{Department of Physics and Astronomy, West Virginia University, WV-26506-6315, Morgantown, USA}
\affiliation{Facultad de Ingenier\'ia-BUAP, Apartado Postal J-39, Puebla, Pue. 72570, Mexico}
\author{M.J. Verstraete}
\affiliation{Nanomat/QMAT/CESAM and European Theoretical Spectroscopy Facility, Li\`ege Universit\'e, B-4000 Sart Tilman, Li\`ege, Belgium}
\author{V. Dmitriev}
\affiliation{Swiss-Norwegian Beam lines, ESRF, BP 220, F-38043 Grenoble Cedex, France}
\affiliation{IRC Smart Materials, Southern Federal University, Zorge Street 5, 344090 Rostov-on-Don, Russia}
\author{A. Bosak}
\affiliation{European Synchrotron Radiation Facility, BP 220, F-38043 Grenoble Cedex, France}

\date{\today}

\begin{abstract}
The high pressure lattice dynamics of rhombohedral antimony have been studied by a combination of diffuse scattering and inelastic x-ray scattering. The evolution of the phonon behavior as function of pressure  was analyzed by means of two theoretical approaches: density functional perturbation theory and symmetry-based phenomenological phase transition analysis. 
This paper focuses on the first structural phase transition, SbI-SbIV, and the role of vibrations in leading the transition. The phonon dispersion exhibits complex behaviour as one approaches the structural transition, with the branches, corresponding to the two transitions happening at high pressure in the Va elements (A7-to-BCC and A7-to-PC) both showing softening.
\end{abstract}

\pacs{00000}


\maketitle


\section{I. INTRODUCTION}

Antimony, like the other group Va elements arsenic, bismuth and pressurized phosphorus, crystallizes under ambient conditions in an A7 rhombohedral structure.
Under pressure it undergoes a sequence of structural phase transitions, adopting high pressure structures, which in many cases are common to those of other group Va elements (for a review see Ref. ~\cite{Katzke2008}), some of which have a surprising complexity as elemental solid structures.
At the highest pressures, all Va elements, excluding molecular nitrogen, adopt a BCC structure. 
While the sequential details vary within the group, what remains unclear in all cases is if the lattice dynamics p a significant role in driving the transitions through these complex structures. 

\begin{figure}
\includegraphics[width=0.40\textwidth]{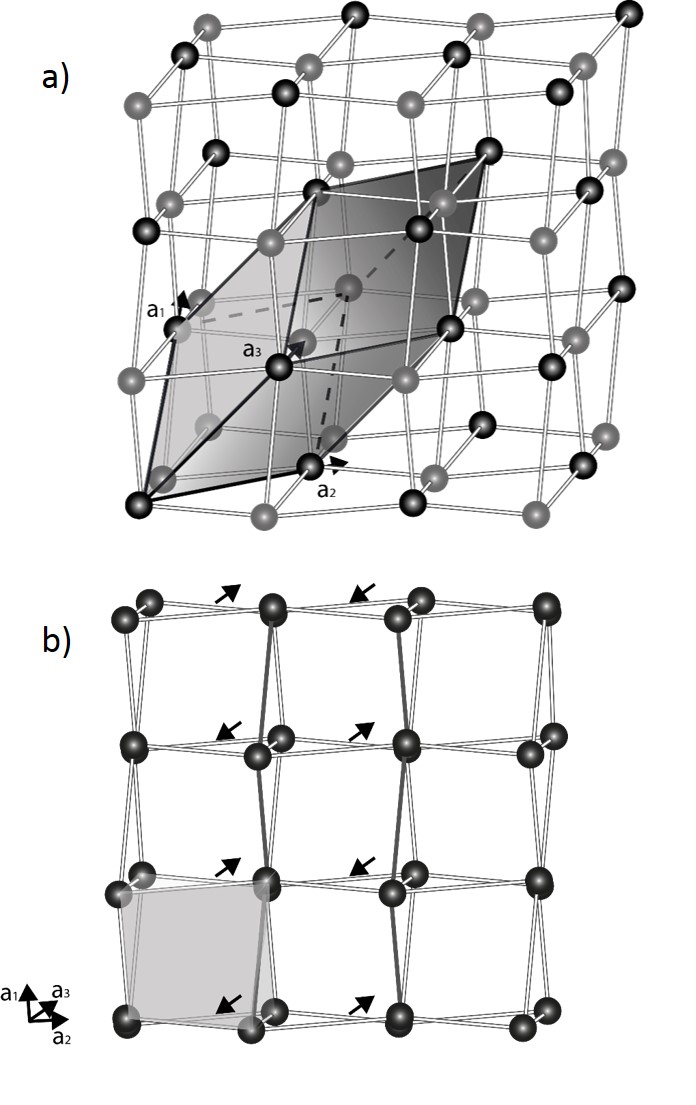}
\caption{a) The A7 structure, where the gray and black balls represent the two interpenetrating FCC (Face centered cubic) lattices and the shaded region corresponds to the rhombohedral unit cell. b) The chains displacement in the A7 structure due to its intrinsic distortions. The black arrows show the direction of the displacement with respect to the PC structure. The shaded region corresponds to the pseudo-cubic cell. }

\label{Rhomb}
\end{figure}

More specifically, in the case of elemental antimony the ambient pressure A7 structure (R$\overline{3}$m) known as SbI, emerges from a simple primitive cubic (PC) structure subjected to two displacements, shown in figure \ref{Rhomb}a. Firstly, an internal-displacement of the two inter-penetrating cubic lattices along a common axis, and secondly, a stretching on the 3-fold axis.
The above mechanism lowers the energy of the system while doubling the unit cell size and opening an energy gap in the electronic density of states, hence the rhombohedral distortion of SbI is often referred as Peierls distortion.

 
Early x-ray diffraction experiments showed that the Peierls distortion decreases at high pressure ~\cite{Iwasaki1986}, in agreement with \textit{ab initio} total energy calculations~\cite{Chang1986,Shick1999}.
However, the simple cubic structure, even though continuously approached under pressure, is never stabilized in Sb.  
Instead, at $\sim$10 GPa antimony adopts a host-guest structure (SbII) where both host and guest have body centered tetragonal lattices (BCT) but the latter is incommensurate with the former along the tetragonal \textit{c}-axis ~\cite{McMahon2000,Schwarz2003}. 
At higher pressures ($\sim$28 GPa), a first-order phase transition takes place towards a BCC structure (SbIII), which is stable up to at least 43 GPa ~\cite{Aoki1983}.
The phase diagram of antimony is further enriched by a second incommensurate monoclinic host-guest structure, known as SbIV, which appears in the narrow pressure range from 8 to 9 GPa~\cite{Degtyareva2004b}.  

Clear fingerprints of the above transitions have been identified in the high pressure Raman spectra of antimony. 
Continuous softening of the zone-center optical phonons was observed over the entire stability range of SbI up to the first transition towards the SbIV phase~\cite{Wang2006,Olijnyk2007,Degtyareva2007}. 
This effect has been observed in other A7 Va elements, such as As and Bi~\cite{Beister1990,Olijnyk2007}, and has been linked to the pressure-induced reduction of the Peierls-like distortion~\cite{Chang1986,Degtyareva2007}.
In addition to this pronounced softening, a large broadening of the Raman peaks was found under pressure, in agreement with \textit{ab initio} calculations~\cite{Serrano2008}. Two potential contributors have been suggested: anharmonic decay channels and a pressure-induced changes in the electron-phonon coupling.

While these investigations have demonstrated the sensitivity of the zone center phonons to both structural and electronic changes under pressure, the role ped by lattice excitations in driving the phase transformations remains unclear. 
In particular, the phononic contribution in the Peierls distortion reduction and in the ultimate stabilization of the BCC structure needs to be clarified. 
Momentum-resolved phonon measurements focused on the A7 to PC and A7 to BCC phase transitions are needed to resolve this issue.
Phonon dispersion curves along certain high symmetry directions have been obtained via earlier inelastic neutron scattering experiments~\cite{Sharp}.
Nevertheless, up to date a pressure-dependent, momentum-resolved lattice dynamical study has not been reported. 
Here we address this open issue through a comprehensive study, combining x-ray diffuse scattering (DS) and high pressure inelastic x-ray scattering (IXS), in conjunction with  \textit{ab initio} calculations and a phenomenological analysis of the displacive transformation.

\section{II. EXPERIMENTAL METHODS}

Commercially available, high quality, single crystals of antimony were obtained from SurfaceNet GMBH ~\cite{Surfacenet}.
A platelet of $\sim$50 $\mu$m thickness was used for ambient pressure measurements while cylinders of $\sim$150 $\mu$m diameter and $\sim$50 $\mu$m thickness were laser cut for the high pressure study.
In both cases the samples were etched using a HCl/HNO$_3$ mixture to remove surface damages.
Rocking curves of  $\Delta \theta \backsim$ 0.1$^\circ$ confirmed the high crystalline quality of the samples.

High pressures were obtained using a gasketed diamond anvil cell (DAC). 
All measurements were performed at room temperature with compressed helium used as a pressure transmitting medium and the pressure calibration was performed using the ruby luminescence ~\cite{Syassen2008}. 
The samples were oriented with the rhombohedral [111] axis (hexagonal $c$-axis) perpendicular to the diamond facets. 
Both crystallographic settings will be used to describe the SbI structure, denoted as HKL$_R$ and HKL$_H$ for the rhombohedral and hexagonal systems respectively.

The DS and IXS experiments were conducted at beamlines ID23 and ID28 of the European Synchrotron Radiation Facility (ESRF) respectively.
A monochromatic beam of with $\lambda$=0.689 \AA, corresponding to 17.994 keV of energy, was used for all the DS measurements. 
The sample was rotated through 360$^\circ$ orthogonal to the incoming beam and diffuse scattering frames were collected with an angular slicing of 0.1$^\circ$. 
A Pilatus 6M (Dectris) detector with a pixel size of 172 x 172 $\mu$m$^2$ was used in single photon counting mode \cite{Pilatus}.  
The CrysAlis software package was used to obtain the orientation matrix and to perform a preliminary data evaluation~\cite{Crysalis}.
All IXS measurements were performed using a 17.794 keV beam (with $\lambda$=0.06968 \AA), corresponding to an energy resolution of 3 meV, 
with a 15 x 13 $\mu$m$^2$ spot size on the sample surface. 
Further details about the experimental setup can be found in reference ~\cite{Krisch2007}.

Additional experiments on a single crystal bismuth sample (SurfaceNet GMBH ~\cite{Surfacenet}) were performed using Pilatus 300K with CdTe sensor material at ID15 ESRF beamline, employing 0.178 \AA~ wavelength and 0.1$^\circ$ angular slicing.

\begin{figure}
\includegraphics[width=\linewidth]{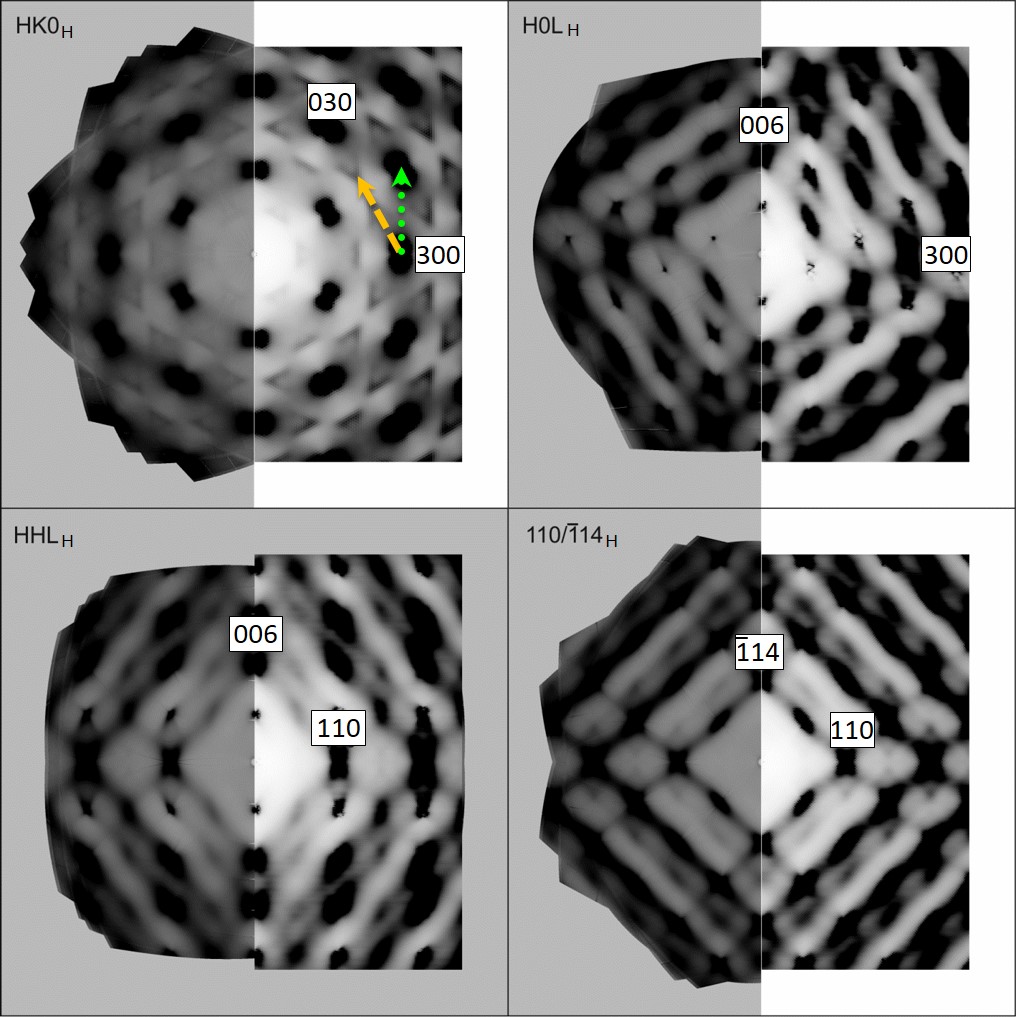}
\caption{Diffuse scattering maps of antimony at ambient condition: the calculated map (right) and the experimental reconstructed planes (left). The yellow (dashed line) and green (dotted line) arrows correspond to two of the diffuse features: the v-shape and diffuse planes, respectively. The hexagonal setting is chosen.} 
\label{DSmaps}
\end{figure} 

\begin{figure*}
\includegraphics[width=\linewidth,]{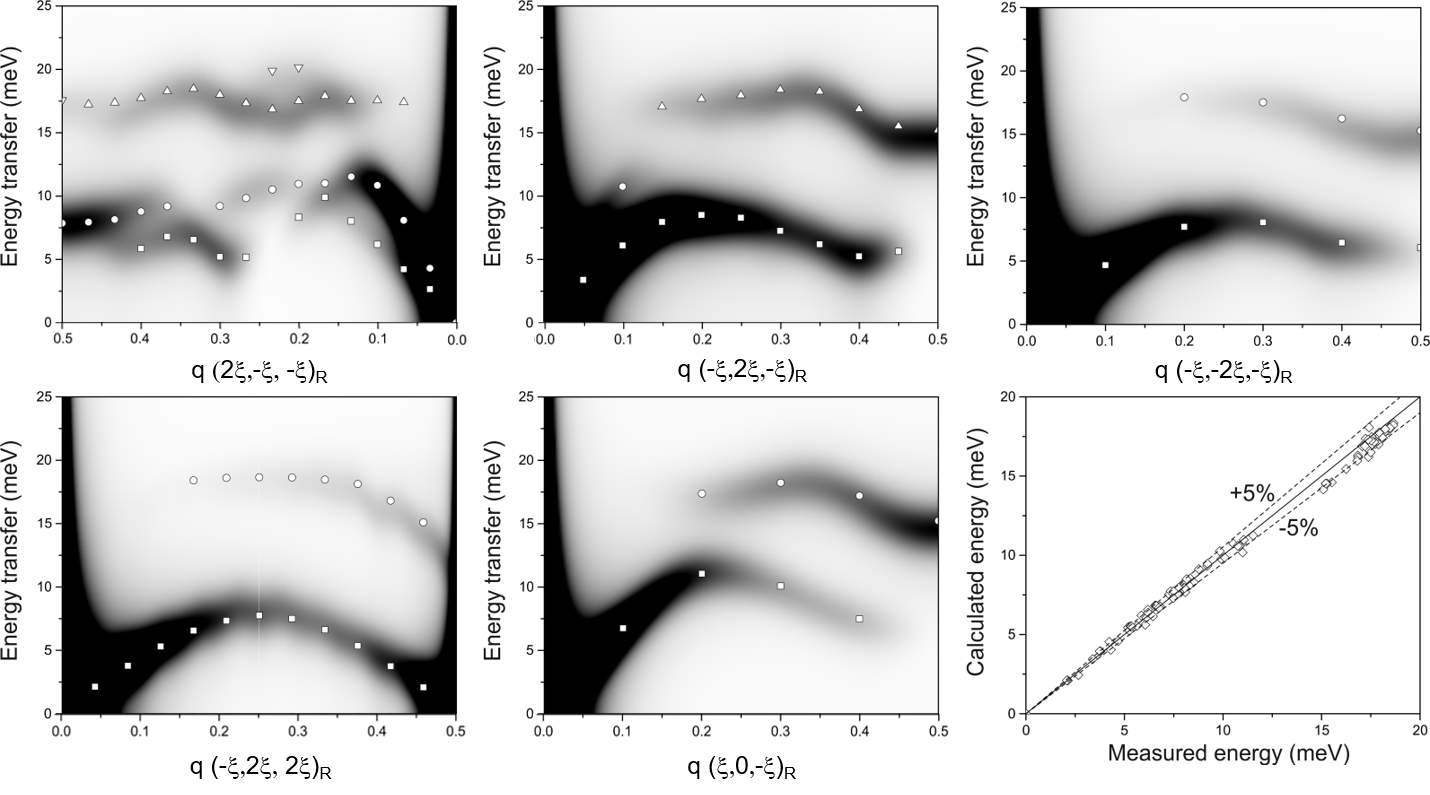}
\caption{Phonon dispersions in different directions of the reciprocal space at ambient conditions. Empty symbols correspond to the fitted frequencies; number of observable branches from the calculation is taken into account. Grayscale map corresponds to the theoretical $S$(\textbf{Q},$E$) convoluted to the realistic resolution function. The deviation between the calculated and measured phonon energy shows a good agreement in the order of a $\pm 5 \%$ error.}
\label{IXs-DS-RT}
\end{figure*}

\section{III. \textit{Ab-initio} CALCULATIONS}

Density functional theory \textit{ab initio} calculations of the lattice dynamics were performed in order to complement and interpret the experiments using the ABINIT software package~\cite{gonze2002first, gonze2016recent, gonze2009abinit, gonze2005brief}.
The antimony electronic structure in the A7 phase was calculated using the Fritz-Haber pseudopotentials by following the description in 
reference~\cite{fuchs1999ab}\footnote{The readers can ask to the authors for the pseudopotentials used in this work}. A grid of 8x8x8 \textbf{k}-points was used to integrate quantities within the Brillouin zone and obtain any electronic property. An energy cutoff of 50 Ha was employed to guarantee energy convergence to less than 10$^{-5}$ eV/atom and stresses less than 10$^{-3}$ GPa. The local density approximation was applied to the exchange and correlation energy~\cite{kohn1965self} and the preconditioning method discussed in 
Ref.~\cite{anglade2008preconditioning} was used to improve the density convergence. The crystal cell optimization and the interatomic force minimization were performed by following the conjugate gradient algorithm as described in Ref.~\cite{gonze1997first}. The atomic coordinates, lattice vectors, and lattice parameters were fully relaxed under the constraint of constant pressures between 0 and 7 GPa. Additional calculations were performed at a constant volume of 53.5 \AA , in order to compare the quality of the calculations with Ref.~\cite{Serrano2008}.

The calculated maps are given by direct calculation in one-phonon scattering where the harmonic approximation is considered. Phonon dispersion relations along selected high symmetry directions were obtained by interpolation of the dynamical matrices corresponding to a \textbf{q}-mesh grid of 8x8x8 at each pressure point. These matrices were calculated within the framework of density functional perturbation theory and the linear response method~\cite{gonze1997dynamical, baroni2001phonons}.

\section{V. RESULTS AND DISCUSSION}


\subsection{a. Ambient pressure}

\begin{figure*}
\includegraphics[width=\linewidth,]{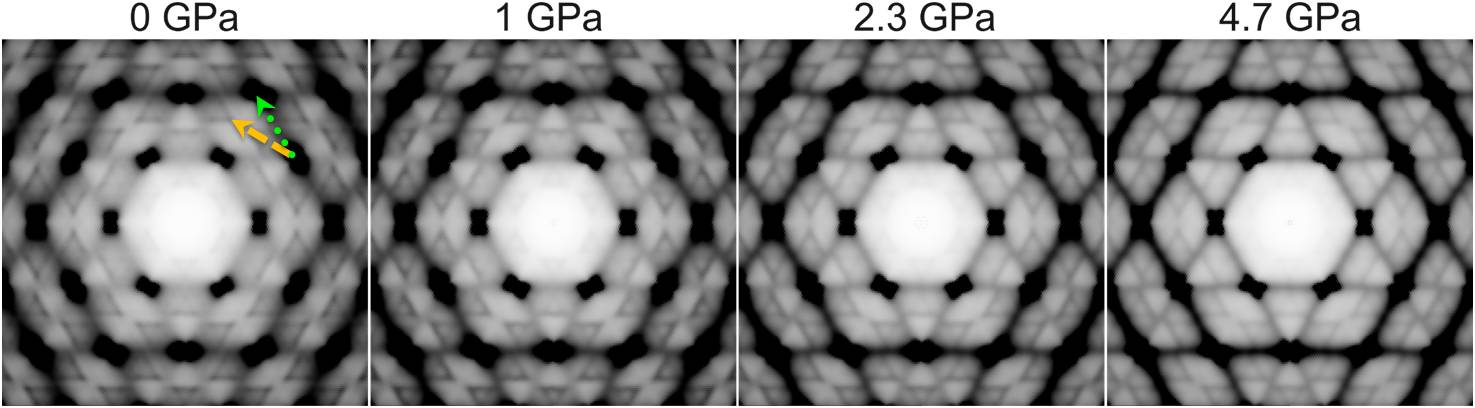}
\caption{HK0$_{H}$ or $<1\overline{1}0/11\overline{2}>_{R}$ maps reconstructed from \textit{ab initio} calculations. Applying pressure, diffuse features in $<\overline{1}10>_{R}$ directions become stronger. The dashed and dotted line (yellow and green in the online version) arrows correspond to the two directions analyzed by IXS: $[\overline{1}2\overline{1}]_{R}$ and $[\overline{1}10]_{R}$, respectively.} 
\label{DScalc}
\end{figure*}

\begin{figure*}
\includegraphics[width=\textwidth]{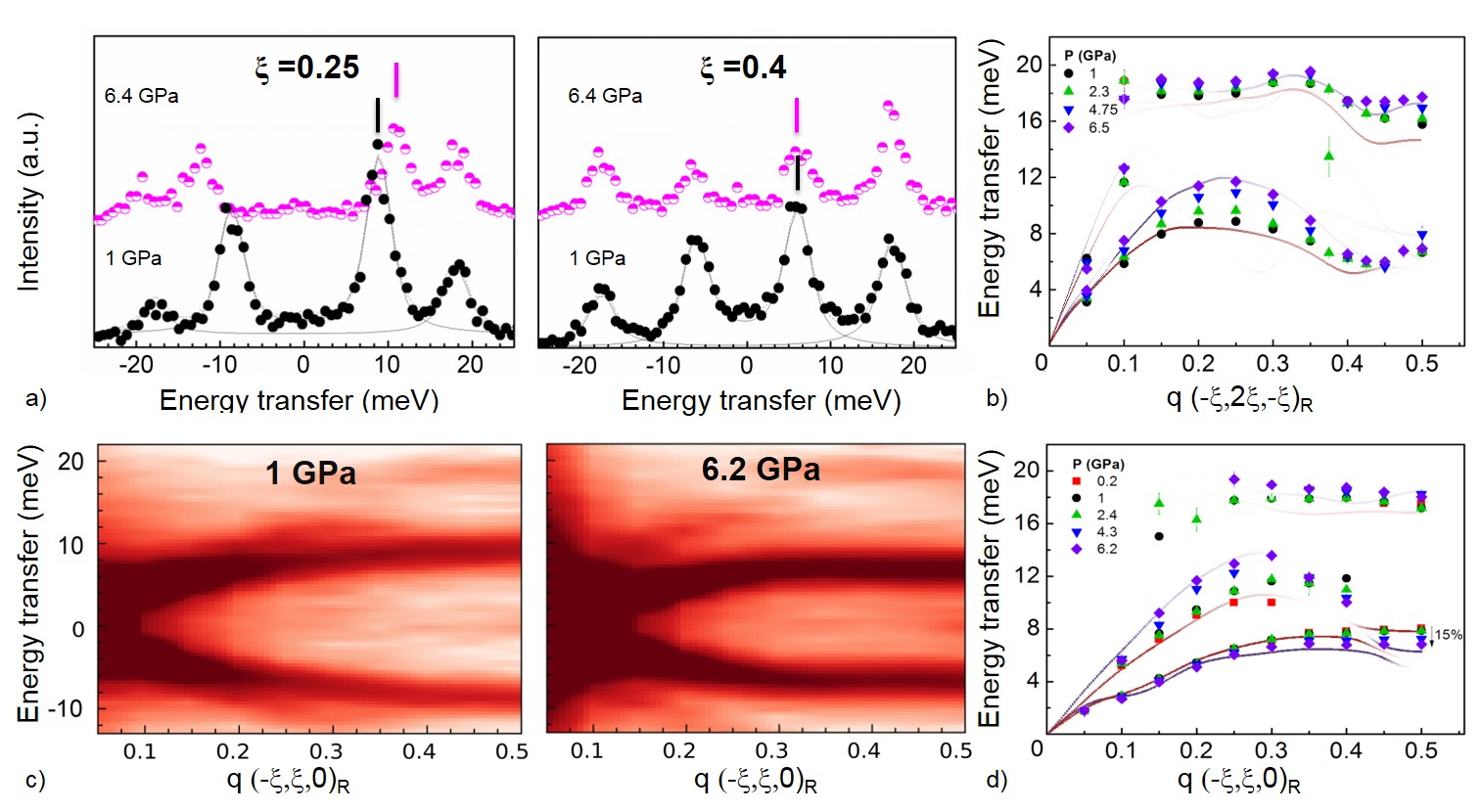}
\caption{ a) From left to right: IXS scans at $\xi$=0.25, on the left, and $\xi$=0.4, on the right, for pressures 1 GPa and 6.5 GPa, in which the lines on the top of the peaks help to visualize the phonon energy obtained from the fitting. b) The fitting results derived from the IXS scans show the general phonon dispersion approaching the first transition. Each symbol corresponds to a specific experimental pressure (color online). The results of the calculations are shows as lines for ambient pressure and 6.5 GPa. The line intensity is proportional to the scattering factor, which was considered for the specific direction. c) Experimental phonon dispersions in the direction $[ \overline{1} 1 0 ]_{R}$ at different applied pressures: 1 GPa, on the left, and 6.2 GPa, on the right; d) the fitting results derived from IXS scans at each applied pressure from 0.2 GPa to 6.2 GPa, compared to the calculations.}
\label{phonon}
\end{figure*}



Figure \ref{DSmaps} shows the experimental diffuse scattering patterns of antimony, at ambient conditions as reconstructed reciprocal space planes compared with the calculated maps. From the plethora of complex diffuse patterns present in the maps of figure \ref{DSmaps}, we focus on those most significant to the approaching phase transition. The first notable feature is the diffuse planes in the $<110/\overline{1}14>_{H}$ map, that resembles 4-fold axis symmetry. The planes are identified by the  $<110>_{R}$ direction normal to the plane or by the $<100>_{PC}$ direction of the pseudo-cubic structure. They seem to be connected to the chains displacement present as intrinsic distortion in the A7 structure, shown in the figure \ref{Rhomb}b in which the black arrows help to visualize the displacement with respect to the non-distorted PC structure. Other notable features  in the HK0$_{H}$ plane: (1) lines in the $<120>_{H}$ direction from the diffuse planes that intersect the HK0$_{H}$; (2) v-shapes in the $<\overline{1}10>_{H}$ direction, shown in figure \ref{DSmaps}.

The remarkable agreement between the experimental and calculated maps adds considerable credence to the accuracy of the calculations. However, the experimental intensities on the map are not resolved in energy, showing just a particular sensitivity for low energy features. For this reason, IXS measurements at ambient conditions were performed in order to demonstrate and confirm the liability of the $ab-initio$ calculations. Figure \ref{IXs-DS-RT} summarizes the ambient pressure IXS dispersion relations showing a discrepancy of less than $ \pm 5 \%$ observed between experiment and theory results. 

Focusing on the phonon dispersions, most directions show a low energy phonon near the zone boundary. One example is in the $[\overline{1}20\overline{1}]_R$ direction, that corresponds to the v-shape diffuse features. In order to further investigate the nature of such diffuse components which have a dynamic nature, we turned to pressure dependent IXS measurements. \\

\subsection{b. High pressure}


The calculated maps of the HK0$_H$ plane at pressures up to 4.7 GPa are shown in figure \ref{DScalc}.  The diffuse planes visible in the $<\overline{1}20>_H$ directions are increasing in intensity under pressure. On the contrary, the v-shaped patterns appear not to show any pressure dependence. Figure \ref{phonon} summarizes the high pressure IXS results along the two directions depicted by green and yellow arrows in figure \ref{DScalc}. \\

In figure \ref{phonon}b, the phonon dispersion curves in the direction $[ \overline{1} 2 \overline{1} ]_{R}$ ($[\overline{1}10]_H$) disp the general trend of phonon hardening with increasing pressure.  The IXS scans at $\xi$=0.25 at 1 GPa and 6.5 GPa provide an example of this behaviour, fig. \ref{phonon}a. The optical phonon also shows hardening across the BZ, an effect which is enhanced at the zone boundary. However a softening, or a "relative softening", meaning an absence of hardening, is observed for the acoustic branch near the BZ boundary. This is shown in detail in the 1 and 6.5 GPa scans at $\xi$=0.4 where the acoustic phonon is unchanged in energy ($\sim$6 meV), see vertical lines in the figure \ref{phonon}a on the right.\\


In the $[\overline{1}10]_{R}$ ($[\overline{1}20]_H$) direction, the LA (longitudinal acoustic) and TA (transverse acoustic) phonons behave differently under pressure, fig. \ref{phonon}c,d. According to the theory there should be an anticrossing between LA and TA around $\xi$=0.43. The intensity of the longitudinal phonon almost disappears while the transverse one remains measurable.  The LA softens by $ \backsim 7 \%$, whereas the TA behaves similarly to the phonon in the $[ \overline{1} 2 \overline{1} ]_{R}$ direction: from $\Gamma$ to $\xi$=0.3 there is a pressure-induced hardening and post $\xi$=0.4 until the zone boundary a softening of $ \backsim 15 \%$, fig. \ref{phonon}d.

While the calculations are generally in good agreement with the experimental results, two main differences are found. Firstly, the calculated optical phonons in the $[ \overline{1} 2 \overline{1} ]_{R}$ direction indicate a greater pressure induced hardening and a shift to lower energies than what was observed experimentally. Secondly, theory predicts a greater softening of the TA than the one observed in the IXS experiment ($\backsim 20 \%$ compared to the 15 $\%$).\\

\begin{figure}
\includegraphics[width=0.75\linewidth,]{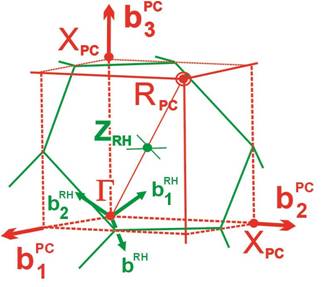}
\caption{Embedment of A7 rhombohedral structure (RH) Brillouin zone, in green, to the cubic (pseudo-PC) one, in red: only the irreducible octant is shown.  R$_{PC}$ is the PC-RH transformation critical point.}
\label{A7-PC}
\end{figure}

\begin{figure}
\includegraphics[width=0.5\textwidth]{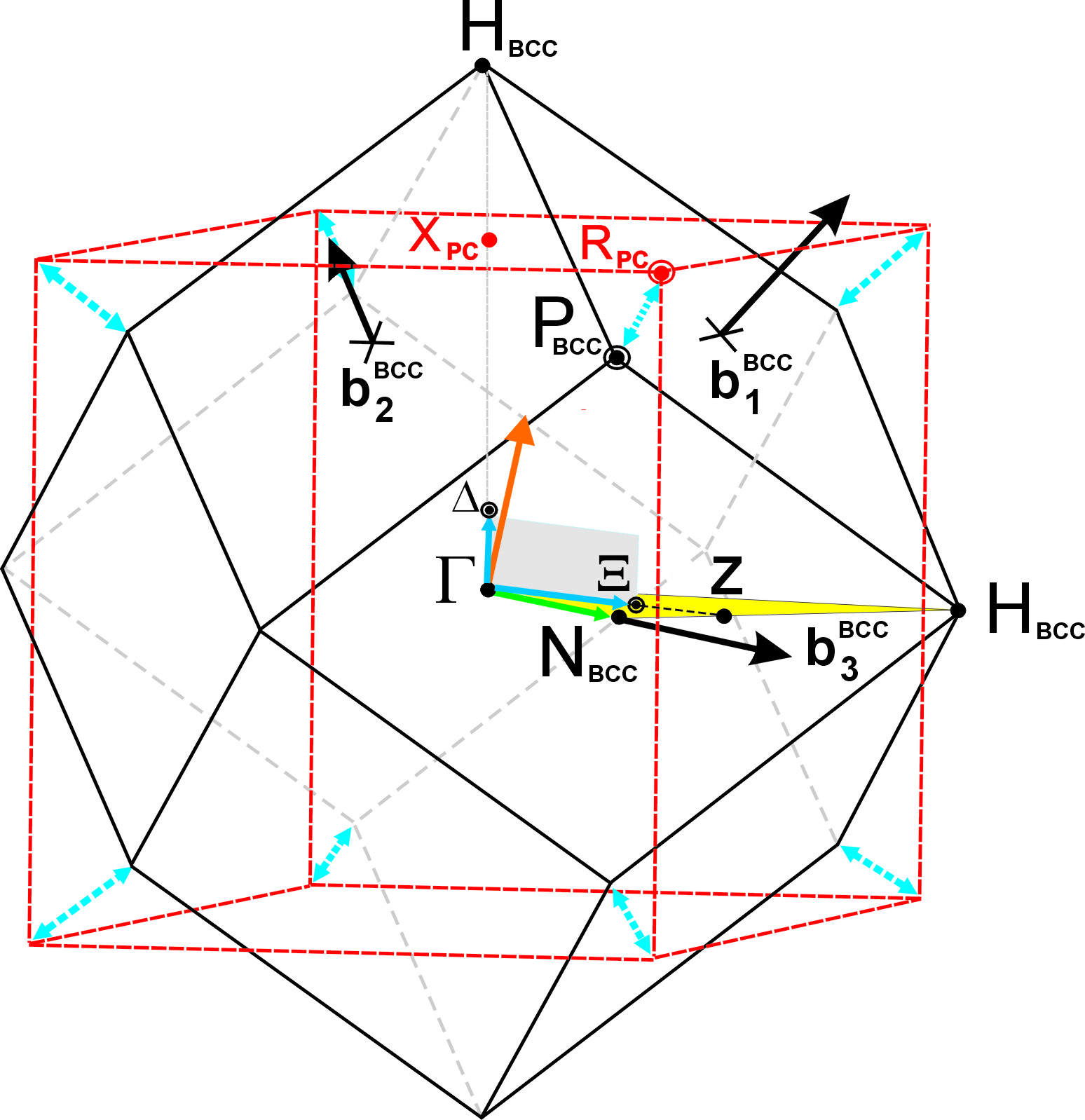}
\caption{ Brillouin zones of the symmetrized A7 rhombohedral structure, as cubic, and high pressure BCC with the high symmetry points of BCC BZ (continuous black line), and that of the primitive cubic BZ (dashed red line). The $<110>_R$ phonon dispersion direction coincides with the $\Gamma$-N$_{BCC}$ line, the orange arrow shows the $<121>_R$ phonon dispersion direction.  All the critical vectors related to the complex phase transitions lie on the yellow plane $\Gamma$-N$_{BCC}$-H$_{BCC}$, with the exception of the vector \textbf{k$_8$}. The two blue arrows corresponds to the directions of the critical vectors for the transition SbIII - SbII (SbIV). Dashed double blue arrows indicate a secondary macroscopic BCC-PC lattice distortion (value $\sim$10 $\% $).  The figure in color is available online.} 

\label{BZ}
\end{figure}

\begin{table*}
\centering
\caption{Critical vectors for principal phase transitions}

\begin{tabular}{ccc}
\hline \hline \\
\large {\textit{Phase transition}} &	\large {\textit{Critical vector}} (k)  & \large {\textit{Point or direction in the BZ}} \footnotemark \\ [0.5ex]
\hline \hline \\
& \textbf{k}$_1^{BCC}$=2/5 (\textbf{b$^{BCC}_1$}-\textbf{b$^{BCC}_2$)}-1/5 \textbf{b$^{BCC}_3$}  & $\Gamma -\Xi -Z$   \\ 
SbIII - SbII (SbIV)	& and  & and  \\ 
&  \textbf{k}$_8^{BCC}$=$\mu$ 2/5 (\textbf{b$^{BCC}_1$}-\textbf{b$^{BCC}_2$}- \textbf{b$^{BCC}_3$}) &  $\Gamma -\Delta -H_{BCC}$  \\
& & \\ 
PC - SbI	& \textbf{k}$_{13}^{PC}$=1/2 (\textbf{b$^{PC}_1$}+\textbf{b$^{PC}_2$}+\textbf{b$^{PC}_3$}) & R$_{PC}$  \\ 
& & \\
SbIII - PC 	& \textbf{k}$_{12}^{BCC}$=1/2 (\textbf{b$^{BCC}_1$}+\textbf{b$^{BCC}_2$}-\textbf{b$^{BCC}_3$})& H$_{BCC}$ \\ [0.5ex] 
\hline \hline \\
\end{tabular}
\label{vector}

a) Direction in the Brillouin zones referred to fig. \ref{BZ}.\\ 

\end{table*} 

\subsection{c. Phase transitions mechanisms and reciprocal space critical vectors}

In order to understand and classify the dynamical behaviour at high pressure, we analyze the corresponding phase transition mechanisms using and completing existing data \cite{Katzke2008}. To this purpose we integrate all the structural data in a unified scheme of a single parent structure, not in \textit{direct} but in \textit{reciprocal} space. As already mentioned in the introduction, the logical choice of a parent structure for the group Va elements is the high-pressure BCC [SbIII, space group Im$\overline{3}$m(Z$^P_{III}$=1), A2 structure type], as it is present in the P-T phase diagrams of all elements belonging to the group. Another structure common to all Va group elements is the low pressure rhombohedral R$\overline{3}$m(Z$^P_{I}$=2) (SbI, A7) phase. However, SbI has been shown to be a slightly distorted version of the simple cubic structure [(pseudo)SbI, Pm$\overline{3}$m(Z$_{pI}$=1)] stabilized at higher pressure in some of the group Va elements (for instance, PVI or AsII) \cite{Degtyareva2004a,Degtyareva2010}. Since SbI under pressure relaxes toward PC and this last structure is reached by other Va group elements, such as arsenic and phosphorus, the phase transition between the two has been included in the discussion. The corresponding critical vector for the A7 to PC transition, \textbf{k}$_{13}$=1/2(\textbf{b}$_1$+\textbf{b}$_2$+\textbf{b}$_3$)$_P$, ends at the R-point of the PC Brillouin zone (BZ), see figure \ref{A7-PC}. In order to simplify the discussion, the pseudo-cubic structure will substitute the distorted A7 for the following part. Thus, without loss of generality, one may consider the quartet of phases `` (pseudo)SbI - SbIV -  SbII - SbIII ``  as representative not only for Sb but for other Va group elements. Such a scheme becomes even better justified if one notes that the intermediate, incommensurate, tetragonal and monoclinic structures (SbII and SbIV) are very similar in all studied elements \cite{Degtyareva2004a,Degtyareva2004b} and structurally interlinked by negligible distortions.

Analysing the ferrodistorsive (Z$_{pI}$=Z$_{III}$) transformation between the pseudo-PC and SbIII one finds that a “group-subgroup” relationship is broken for their space groups Pm$\overline{3}$m(Z$_{pI}$=1) and Im$\overline{3}$m(Z$_{III}$=1). The latter conclusion is also valid for the R$\overline{3}m(2)$ - Im$\overline{3}$m(1) transformation. A phenomenological scheme for such \textit{reconstructive} phase transitions should contain an intermediate phase whose space group is a common subgroup, normally maximal, for both high-symmetry phases \cite{Toledano1996}. A relevant geometrically optimal pathway for Sb would occur via antiparallel shifts of alternating (001)$_{BCC}$ atomic planes in the $\pm$[110]$_{BCC}$ directions. The shifts reduce the crystal symmetry from cubic to orthorhombic Cmcm(2) for general magnitude displacements, then restore it to cubic for special shifts. The corresponding dynamic instability ( `` soft mode ``) develops at the H$_{BCC}$-point [\textbf{k}$_{12}^{BCC}$=1/2(\textbf{b}$_1^{BCC}$+\textbf{b}$_2^{BCC}$-\textbf{b}$_3^{BCC}$)], see figure \ref{BZ}. However, energy minimization criteria, due to the deviation of the atomic interaction character from the simplest spherical, can produce an alternative pathway and this is the case in Sb. It has been shown in earlier DFT calculations \cite{Haussermann2002}, that the complex interp between electrostatic (Madelung) energy E$_{Mad}$ and the band energy E$_{band}$ stabilizes the pressure controlled structural sequence in Sb: E$_{band}$ governs the structural stability at low pressure while E$_{Mad}$ dominates at high pressure, stabilizing the densely packed BCC, giving rise to the complex intermediate, incommensurate, structures and directing atoms along a non-trivial pathway from PC to BCC.

Moreover, a universal trend to stabilize intermediate, incommensurate, structures in the vicinity of reconstructive phase transitions was recently predicted in the framework of more general approach \cite{Korzhenevskii2015}. A model-free, symmetry based theory introduces a universal mechanism for the formation and stabilisation of such inhomogeneous states. The mechanism accounts for elastic properties of the crystal lattice via a bi-linear coupling between critical displacement gradients (possible non-uniform order-parameter distribution) and secondary strains. The importance of the latter in reconstructive mechanisms is widely recognized.

Propagation of distortion in the intermediate SbII and SbIV structures is defined by two reciprocal space vectors: (i) the long-periodic but commensurate \textbf{k}$_1$=2/5(\textbf{b}$_1$-\textbf{b}$_2$)-1/5\textbf{b}$_3$ which lies in the N$_{BCC}$-$\Gamma$ -H$_{BCC}$ plane, and (ii) the incommensurate \textbf{k}$_8$=$\mu$(\textbf{b}$_1$+\textbf{b}$_2$-\textbf{b}$_3$), which belongs to the $\Gamma$ -$\Delta$ -H$_{BCC}$ line perpendicular to this plane. Remarkably, the k$_8^{BCC}$ varies towards the H$_{BCC}$-point, and BCC (SbIII), therefore, could be considered as a ''lock-in'' phase or vice versa. Figure \ref{BZ} and table \ref{vector} summarise the set of critical points, lines and planes in the BCC reciprocal space.

\subsection{d. Phase stability}

The measurements presented in this work, together with previously reported experimental data \citep{Degtyareva2007,Wang2006},  allow us to conclude that antimony follows the general trend of group Va elements: the rhombohedral A7 structure seeks to increase its symmetry and stabilize the PC phase. The two main observations supporting this conclusion are as follow. First, the Raman results on SbI \cite{Degtyareva2007,Wang2006} revealed simultaneous softening of two modes, $\Gamma_1^{+}$ (A$_g$) and $\Gamma_3^{+}$(E$_g$), as one increases pressure. These modes originate from the BZ folding which transfers the critical, triply degenerated, R$_5^{+}$ phonon mode from the PC BZ R-node to the A7 BZ centre ($\Gamma$-point, Fig. \ref{A7-PC} ), splitting it into two Raman active modes $\Gamma_1^{+}$ and $\Gamma_3^{+}$. Second, the phonon dispersion branches observed by IXS in the $<121>_R$ direction, highlighted as orange arrow in figure \ref{BZ}. It lies near the $\Gamma R$ direction and exhibits a ''relative softening'' under pressure. The partial mode softening seen in Raman study \cite{Degtyareva2007,Wang2006} combined with the current results and analysis indicate one is approaching the stability limit of the rhombohedral phase with respect to cubic one. In both cases, only partial softening is observed. 

Instead, antimony follows a pathway containing monoclinic and tetragonal incommensurate phases that is favoured energetically prior to stabilise the BCC phase. To quantify this, we measured by IXS the $<110>_R$ direction (green arrow in figure \ref{BZ}). It is near the BCC-SbII critical vector and in plane with the BCC-pseudo-PC(A7) critical vector. A more intense softening is found, with a decrease in energy of the $\sim$ 15 $\%$.   

\begin{figure}
\includegraphics[width=\linewidth]{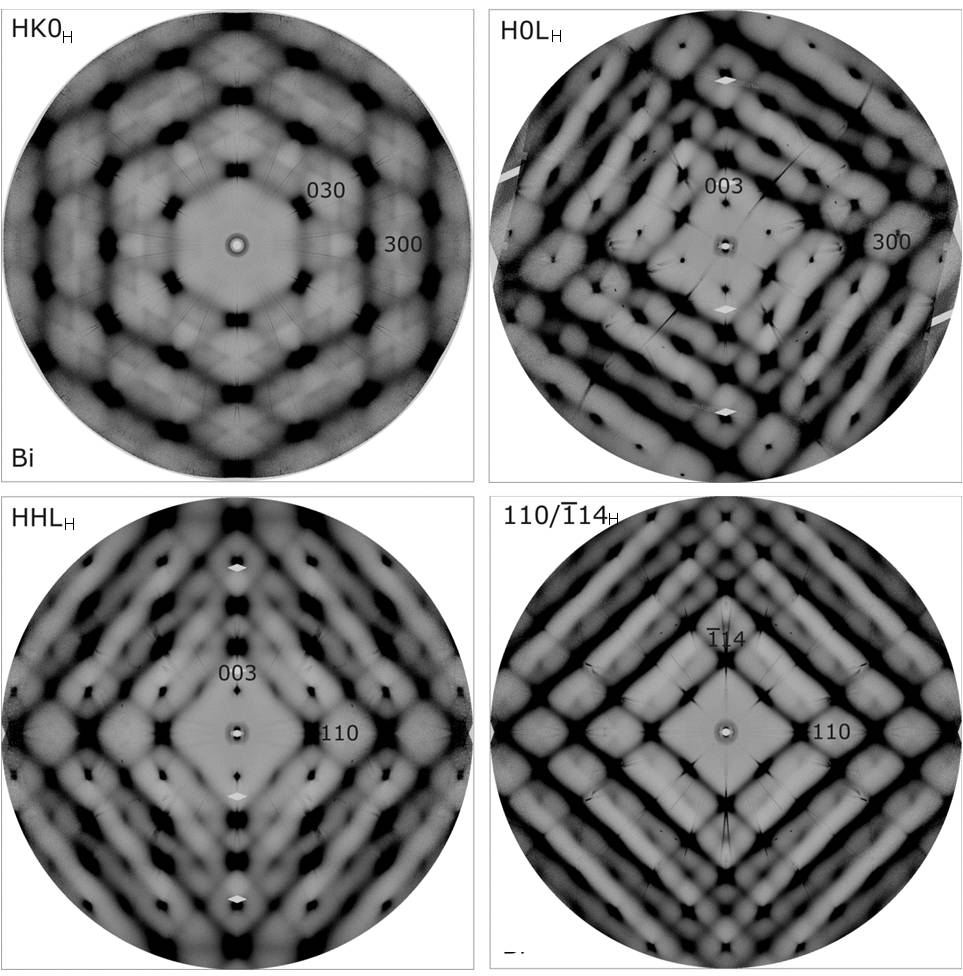}
\caption{Diffuse scattering maps of bismuth at ambient conditions. In this case, the entire maps are reconstructed from experimental results. The hexagonal setting is chosen to classify the planes.} 
\label{Bi}
\end{figure}

We conclude our work by reporting the diffuse scattering maps collected on other element of the Va group, bismuth. In figure \ref{Bi}, the measured maps did not reveal any qualitative features different from those of antimony, therefore, we expect a similar pressure trend for this element. Indeed, according to earlier studies, bismuth follows a similar structural path to antimony \cite{Katzke2008,McMahon2000,Degtyareva2004a,Aoki1982}.


\section{Conclusion}

The lattice dynamics of antimony at both ambient and high pressure conditions show anomalies that are understood by means of \textit{ab initio} calculations and phase transition mechanisms analysis. This study focused on the pressure dependence of the phonon dispersions along two critical directions in the reciprocal space corresponding to the canonical Va group phase transitions. The reconstructive phase transition analysis was employed with a focus on those two mechanisms: I) the A7-PC transition, which describes the relaxation of the rhombohedral distorted structure that is never reached in this particular system and II) the A7-BCC transition, which passes through two incommensurate structures. In the first case, the relevant phonon dispersion shows a relative softening under pressure, while for the second one, a bigger pressure-induced softening of $\backsim 15 \%$ is observed along the direction. While both transition mechanisms are mirrored in the phonon dispersions of antimony, the strongest phonon softening is related to A7-BCC transition, which is also the more energetically favorable. The competition of those two mechanisms induces the observed complex behavior of antimony under pressure. \\

This paper is part of the PhD project of A.Minelli. For more information and supplementary data, please consult the online version~\cite{Minelli}.

\section{Acknowledgement}

The authors would like to thank Karl Syassen for the fruitful discussions and encouragement. Denis Gambetti and Jeroen Jacobs are also thanked for technical support during the experiment in ID28 and Sasha Popov for the scientific support in the preliminary diffuse scattering experiment at ID23.
A. H. R. acknowledge the support of NSF under grants 1434897, 1740111 and DOE under grant 
DOE DE-SC0016176. This work used the Extreme Science and Engineering Discovery Environment (XSEDE), supported by National Science Foundation grant number OCI-1053575. Additionally, the authors acknowledge the computing support from Texas Advances Computer Center (TACC) with the Stampede 2 and Bridges supercomputer at Pittsburgh Supercomputer Center.


\bibliography{Sb}
\end{document}